\documentstyle[12pt,epsfig]{article}  
\begin{document} 
\baselineskip=20pt

%%%%%%%%%%%%%%%%%%%%%%%% NEW DEFINITIONS
\def\la{\mathrel{\mathpalette\fun <}}
\def\ga{\mathrel{\mathpalette\fun >}}
\def\fun#1#2{\lower3.6pt\vbox{\baselineskip0pt\lineskip.9pt
\ialign{$\mathsurround=0pt#1\hfil##\hfil$\crcr#2\crcr\sim\crcr}}} 
%%%%%%%%%%%%%%%%%%%%%%%% TITLE PAGE

\begin{titlepage} 
\begin{center}
{\Large \bf Calorimetric transverse energy-energy correlations as a probe of 
jet quenching
} \\

\vspace{4mm}

I.P.~Lokhtin$^a$, 
L.I.~Sarycheva$^b$,  
A.M.~Snigirev$^c$  \\
M.V. Lomonosov Moscow State University, D.V. Skobeltsyn Institute of Nuclear 
Physics, \\
119992, Vorobievy Gory, Moscow, Russia 
\end{center}  

\begin{abstract}  
The sensitivity of calorimetric energy-energy correlation function to the 
medium-induced energy loss of fast partons in high multiplicity heavy ion
interactions is demonstrated at the appropriate selection of events for the 
analysis, namely, the availability of one high-$p_T$ jet in an event at least
and using the procedure of ``thermal'' background subtraction. Without jet 
trigger this correlation function manifests the global structure of 
transverse energy flux: the correlator is isotropic for central collisions, and 
for non-central collisions it is sensitive to the azimuthal anisotropy of 
energy flow reproducing its Fourier harmonics but with the coefficients 
squared.
\end{abstract}

\bigskip

\vspace{100mm}
\noindent
---------------------------------------------\\
$^a$ e-mail: igor@lav01.sinp.msu.ru\\
$^b$ e-mail: lis@alex.sinp.msu.ru\\
$^c$ e-mail: snigirev@lav01.sinp.msu.ru\\
\end{titlepage}   

\section{Introduction} 

The QGP creation and properties of this new type of superdense matter are
extensively discussed in the current literature~\cite{qm02} dedicated to high
energy heavy ion physics. In the last several years many new phenomena have been
observed at RHIC, such as strong elliptic flow~\cite{ackermann} and suppression of the
high transverse momentum two particle back-to-back correlations~\cite{adler}. 
These observations together with other important evidences support the idea that 
a dense partonic matter (QGP) has been created in such high energy nuclear
collisions. 

The center of mass energy for heavy ion collisions at the LHC will exceed that at
RHIC by a factor of about 30. This provides exciting opportunities for
addressing unique physics issues in a new energy domain, where hard and 
semi-hard QCD multi-particle production can certainly dominate over underlying 
soft events~\cite{cern}. The methodological advantage of azimuthal jet 
observables is that one needs to reconstruct only the azimuthal position of the 
jet in heavy ion event, that can be done with fine enough resolution (slightly 
worse as compared with $pp$ case but still lesser than the typical azimuthal 
size of a calorimeter tower~\cite{cern,baur}). However the observation of the 
jet azimuthal anisotropy due to medium-induced partonic energy 
loss~\cite{lokhtin03, lokhtin02} requires event-by-event determination of the 
nuclear event plane. Recently the ability to reconstruct the reaction plane 
using calorimetric measurements has been shown in~\cite{note} as well as the 
possibility to observe jet azimuthal anisotropy without direct determination of 
the event plane, considering the second~\cite{Lokhtin02} and 
higher~\cite{Lokhtin03} order correlators between the azimuthal position of the 
jet axis and the angles of the particles not incorporated in the jet. This 
calorimetric information allows us to study also the energy-energy correlations 
which are sensitive to partonic energy loss (as it has been pointed out 
in~\cite{pan}) and this investigation does not demand also the reconstruction 
of the event plane which is still far from to be completed. 

The observed at RHIC suppression of the high transverse momentum two particle
back-to-back correlations~\cite{adler} supports strongly the believe that the
calorimetric measurements of energy-energy correlations at LHC will be able to
provide a new additional information about medium-induced energy loss of fast
partons. Namely these correlations are the subject of the present investigation.

\section{Energy-energy correlation functions in $e^+e^-$, hadronic and nuclear
collisions}

The energy-energy correlation function $\Sigma$ has been used by all for LEP
experiments \cite{abreu} at CERN and the SLD experiment \cite{abe} at SLAC to 
measure the strong coupling constant $\alpha_s$ in $e^+e^-$-annihilation at 
the $Z^0$ resonance with a high accuracy.
$\Sigma$ is defined as a function of the angle $\chi$ between two particles $i$
and $j$ in the following form 
\begin{equation}
\label{1}
\frac{d\sum(\chi)}{d\cos(\chi)} = \frac{\sigma}{\Delta \cos(\chi) N_{\rm event}}
\sum \limits_{{\rm event}} \sum\limits_{i\not= j} \frac{E_iE_j}{E^2}, 
\end{equation}
where $E$ is the total energy of the event, $E_i$ and $E_j$ are the energies of
the particles $i$ and $j$. The sum runs over all pairs $i$, $j$ with
$\cos(\chi)$ in a bin width $\Delta \cos(\chi)$: 
$$\cos(\chi) - \Delta\cos(\chi)/2 < \cos(\chi) < \cos(\chi) +
\Delta\cos(\chi)/2.$$
$\sigma$ is the total cross section for $e^+e^- \to$ hadrons. The limits
$\Delta\cos(\chi) \to 0$ and $N_{\rm event} \to \infty$ have to be taken in
(\ref{1}).

This function can be calculated in perturbative QCD as a series in $\alpha_s$:
\begin{equation}
\label{2}
\frac{1}{\sigma_0} \frac{d\Sigma(\chi)}{d\cos(\chi)} =
\frac{\alpha_s(\mu)}{2\pi}A(\chi) + \left(\frac{\alpha_s(\mu)}{2\pi}\right)^2
\left(\beta_0 \ln\left(\frac{\mu}{E}\right) A(\chi) + B(\chi)\right) +
O(\alpha_s^3),
\end{equation}
where $\beta_0 = (33-2n_f)/3$, $n_f$ is the number of active flavours at the
energy $E$. The first order term $A(\chi)$ has been calculated by Basham et al.\
\cite{basham} from the well-known one gluon emission diagrams $\gamma^*,Z^0 \to q\bar
qg$ with the result 
\begin{equation}
\label{3}
A(\chi) = C_F(1+\omega)^3\frac{1 + 3\omega}{4\omega} ((2-6\omega^2) \ln(1 +
1/\omega) + 6\omega - 3),
\end{equation}
where $C_F=4/3,~\omega = \cot^2(\chi)/2$ and 
$\chi$ is the angle between any of the three
partons. This allowed one to determine the strong coupling constant directly
from a fit to the energy-energy correlations in $e^+e^-$-annihilation, since
$\Sigma$ is proportional to $\alpha_s$ in the first order. 

In hadronic and nuclear collisions jets are produced by hard scattering of
partons. In this case it is convenient to introduce the transverse energy-energy
correlations which depend very weakly on the structure functions \cite{ali} and
manifest directly the topology of events. Thus, for instance, in high-$p_T$
two-jet events the correlation function is peaked at the azimuthal angle $\varphi =
0^\circ$ and $\varphi = 180^\circ$, while for the 
isotropic background this function is independent of $\varphi$. 

Utilization of hard jet characteristics to investigate QGP in heavy ion 
collisions is complicated because of a huge multiplicity of ''thermal'' 
secondary particles in an event. Various estimations give from 1500 to 6000 
charged particles per rapidity unit in a central Pb--Pb collision at LHC energy, 
and jets can be really reconstructed against the background of energy flux 
beginning from some threshold jet transverse energy $E_T^{\rm jet} \sim 50-100$ 
GeV \cite{cern,baur,kruglov}. The transverse energy-energy correlation function 
is just sensitive to the jet quenching as well as to the global structure of 
energy flux (i.e. its anisotropy for non-central collisions) if we select 
events by an appropriate way. The generalization of (\ref{1}) for calorimetric  
measurements of the energy flow is straightforward: 
\begin{equation}
\label{4}
\frac{d\Sigma_T(\varphi)}{d\varphi} = \frac{1}{\Delta\varphi~ N_{\rm
event}}\sum\limits_{{\rm event}}\sum\limits_i \frac{E_{Ti}
E_{T(i+n)}}{(E_T^{\rm vis})^2},
\end{equation}
where
$$E_T^{\rm vis} = \sum\limits_i E_{Ti}$$
is the total transverse energy in $N$ calorimetric sectors covering the full 
azimuth, $E_{Ti}$ is the transverse energy deposition in a sector $i$
($i=1,...,N$) in the 
considered pseudo-rapidity region $|\eta|$, $n=[\varphi/\Delta\varphi]$ (the 
integer part of the number $\varphi/\Delta\varphi$), $\Delta\varphi \sim 0.1$
is the typical azimuthal size of a calorimetric sector. In the continuous 
limit $\Delta\varphi \to 0$, $N = [2\pi/\Delta\varphi] \to \infty$ the equation
(\ref{4}) reads 
\begin{equation}
\label{Sigma_T}
\frac{d\Sigma_T(\varphi)}{d\varphi} = \frac{1}{N_{\rm
event}}\sum_{{\rm event}}\frac{1}{(E_T^{\rm vis})^2}\int\limits_0^{2\pi}d\phi
\frac{dE_T}{d\phi}(\phi)\frac{dE_T}{d\phi}(\phi + \varphi),
\end{equation}
where
$\displaystyle E_T^{\rm vis} = \int\limits_0^{2\pi} 
d\phi~\frac{dE_T}{d\phi}(\phi)$, 
~~$\displaystyle \frac{dE_T}{d\phi}$ is the distribution of transverse energy flow
over azimuthal
angle $\phi$. 

The events with high-$p_T$ jet production are rare events and their 
contribution to $\Sigma_T$ is negligible without the special jet trigger. As a 
result, the transverse energy-energy correlator is expected to be independent 
of $\varphi$ for central collisions due to the isotropy of energy flow on the 
whole. For non-central collisions with the clearly visible elliptic flow of the 
transverse energy, 
\begin{equation}
\label{5}
\frac{dE_T(\phi)}{d\phi} = \frac{E_{T}^{\rm vis}}{2\pi}[1 + 2v_2\cos(2(\phi - \psi_R))],
\end{equation}
this correlator $d\Sigma_T/d\varphi$ is independent of the reaction plane angle $\psi_R$
and is calculated explicitly:
\begin{equation}
\label{6}
\frac{d\Sigma_T(\varphi)}{d\varphi} = \frac{1}{2\pi}[1 + 2v_2^2 \cos(2\varphi)].
\end{equation}
The strength of  collective flow (\ref{5})
being determined by the value of $v_2$, and the intensity of oscillations in
(\ref{6}) --- by $v_2^2$~! Moreover one can prove that $d\Sigma_T/d\varphi$ reproduces all
Fourier harmonics of transverse energy flow decomposition but with the
coefficients squared.

\section{The model to simulate heavy ion events without and with jets at the LHC} 

In order to see the manifestation of the jet structure we must cut the
background events, i.e.\ consider the sum in (\ref{4}) over only the events
containing at least one jet with $E_T^{\rm jet} > E_T(\rm threshold) \sim 100$ 
GeV and subtract the background from soft thermal particles using its isotropy
on the whole. 
We demonstrate the productivity and effectiveness of such strategy in the
framework of our well worked-out model of jet passing through a medium 
applied early to the calculation of various observables sensitive to the 
partonic energy loss: the impact parameter dependence of jet production
~\cite{lokhtin00}, the mono/dijet rate enhancement~\cite{lokhtin99, bass},
the dijet rate dependence on the angular jet cone~\cite{lokhtin98}, 
the elliptic coefficient of jet azimuthal anisotropy~\cite{lokhtin03,
lokhtin02, note, Lokhtin02, Lokhtin03}, 
the anti-correlation between  softening jet fragmentation
function and suppression of the jet rate~\cite{phl03}. For model details one can
refer to these mentioned papers (mainly,~~\cite{lokhtin00,lokhtin98,phl03}).
Here we note only the main steps essential for the present investigation. 

PYTHIA$\_6.2$~\cite{pythia} was used to generate the initial jet distributions 
in nucleon-nucleon sub-collisions at $\sqrt{s}=5.5$ TeV.
After that event-by-event Monte-Carlo simulation of rescattering and energy 
loss of jet partons in QGP was performed. The approach relies on   
accumulative energy losses, when gluon radiation is associated with each 
scattering in expanding medium together including the interference effect by 
the modified radiation spectrum as a function of decreasing temperature.
Such numerical simulation of free path of a hard jet in QGP allows any 
kinematical characteristic distributions of jets in the final state to be 
obtained. Besides the different scenarios of medium evolution can be considered. 
In each $i$-th scattering a fast parton loses energy 
collisionally and radiatively, $\Delta e_i = t_i/(2m_0) + \omega _i$, 
where the transfer momentum squared $t_i$ is simulated according to the
differential cross section for elastic
scattering of a parton with energy $E$ off the 
``thermal'' partons with energy (or effective mass) $m_0 \sim 3T \ll E$ 
at temperature $T$, and
$\omega _i$ is simulated according to the energy spectrum of coherent
medium-induced gluon radiation in BDMS formalism~\cite{baier}. 
Finally we suppose that in every event the energy of an initial parton 
decreases by the value $\Delta E= \sum _i \Delta e_i$. 

The medium was treated as a boost-invariant longitudinally expanding 
quark-gluon fluid, 
and partons as being produced on a hyper-surface of equal proper times 
$\tau$~\cite{bjorken}. For certainty we used the initial conditions 
for the gluon-dominated plasma formation 
expected for central Pb$-$Pb collisions at LHC~\cite{esk}: 
$\tau_0 \simeq 0.1$ fm/c, $T_0 \simeq 1$ GeV. For non-central collisions we 
suggest the proportionality of the initial energy density to the ratio of 
nuclear overlap function 
and effective transverse area of nuclear overlapping~\cite{lokhtin00}.

The energy-energy correlator depends on not only an absolute value of partonic
energy loss, but also on an angular spectrum of in-medium radiated gluons. Since 
coherent Landau-Pomeranchuk-Migdal radiation induces a strong dependence of the radiative energy 
loss of a jet on the angular cone 
size~\cite{lokhtin98,baier,Zakharov:1999,urs,vitev}, it will soften particle 
energy distributions inside the jet, increase the multiplicity of secondary 
particles, and to a lesser degree, affect the total jet energy. On the other 
hand, collisional 
energy loss turns out to be practically independent of jet cone size and causes 
the loss of total jet energy, because the bulk of ``thermal'' particles knocked out of the 
dense matter by elastic scatterings fly away in almost transverse direction relative to 
the jet axis~\cite{lokhtin98}. Thus although the radiative energy loss of an 
energetic parton dominates over the collisional loss by up to an order of 
magnitude, the relative contribution of collisional loss of a jet growths with 
increasing jet cone size due to essentially different angular structure of loss 
for two mechanisms~\cite{lokhtin98}. Moreover, the total energy loss of a jet 
will be sensitive to the experimental capabilities to detect low-p$_T$ 
particles -- products of soft gluon fragmentation: thresholds for giving signal 
in calorimeters, influence of strong magnetic field, etc.~\cite{baur}. 

Since the full treatment of angular spectrum of emitted gluons is rather 
sophisticated and 
model-dependent~\cite{lokhtin98,baier,Zakharov:1999,urs,vitev}, we considered 
two simple parameterizations of distribution of in-medium radiated gluons over
the emission angle $\theta$. The ``small-angular'' radiation spectrum was
parameterized in the form
\begin{equation} 
\label{sar} 
\frac{dN^g}{d\theta}\propto \sin{\theta} \exp{\left( -\frac{(\theta-\theta
_0)^2}{2\theta_0^2}\right) }~, 
\end{equation}
where $\theta_0 \sim 5^0$ is the typical angle of coherent gluon radiation estimated
in work~\cite{lokhtin98}. The ``broad-angular'' spectrum has the form 
\begin{equation} 
\label{war} 
\frac{dN^g}{d\theta}\propto \frac{1}{\theta}~.  
\end{equation}
We believe that such simplified treatment here is enough to demonstrate
the sensitivity of the energy-energy correlator to
the medium-induced partonic energy loss. 

The following kinematical cuts on the jet transverse energy and pseudo-rapidity 
were applied: $E_T^{\rm jet} > 100$ GeV and $|\eta^{\rm jet}| < 1.5$. After 
this the dijet event is imposed upon the Pb$-$Pb event, which was 
generated using the fast Monte-Carlo simulation 
procedure~\cite{Lokhtin02,kruglov,gener} giving a hadron (charged and neutral 
pion, kaon and proton) spectrum as a superposition of the thermal distribution 
and collective flow. To be definite, we fixed the following ``freeze-out'' 
parameters: the temperature $T_f = 140$ MeV, the collective  longitudinal 
rapidity $Y_L^{max}=3$ and the collective transverse rapidity $Y_T^{max}=1$. We 
set the Poisson multiplicity distribution. For non-central collisions, the 
impact parameter dependence of the multiplicity was taken into account by a 
simple way, just suggesting that the mean multiplicity of particles is 
proportional to the nuclear overlap function. We also suggested~\cite{Lokhtin02} 
that the spatial ellipticity of the ``freeze-out'' region is directly related 
to the initial spatial ellipticity of the nuclear overlap zone. Such ``scaling'' 
allows one to avoid using additional parameters and, at the same time, results 
in an elliptic anisotropy of particle and energy flow due to the dependence of 
the effective transverse size of the ``freeze-out'' region on the azimuthal 
angle of a ``hadronic liquid'' element. Obtained in such a way the azimuthal 
distribution of particles is described well by the elliptic form (\ref{5}) for 
the domain of reasonable impact parameter values.  

\section{Numerical results and discussion}

At first, we became convinced that for non-central collisions with the elliptic
anisotropy of particle and energy flow (generated in the framework of  
simple Monte-Carlo procedure described above) the energy-energy correlation 
function $d\Sigma_T/d\varphi$ (\ref{Sigma_T}) is closely followed the formula 
(\ref{6}). 

Further, only central collisions (in which the effect of jet quenching is
maximum and ``thermal'' background is azimuthally isotropic) will be considered. 
In order to extract jet-like energy-energy correlator from the ``thermal''
background in high multiplicity environment, the energy deposition in each 
calorimeter sector $i$ is recalculated event-by-event as
\begin{equation}  
\label{erecal} 
E_{Ti}^{\rm new} ( \varphi ) = E_{Ti} ( \varphi ) - \overline{E_{Ti}( \varphi )}
-k\cdot D_T~,
\end{equation} 
where $\overline{E_{Ti}( \varphi )}=\frac{1}{N}\sum\limits_i \left( E_{Ti}^{\rm 
thermal}( \varphi )+ E_{Ti}^{\rm jet}( \varphi )\right) $ is the average energy 
deposition in a calorimeter sector and $D_T=\sqrt{\overline{(E_{Ti}( 
\varphi ))^2}-(\overline{E_{Ti}( \varphi )})^2}$ is the energy dispersion in the
given event, $N$ is the total number of calorimeter sectors. If 
$E_{Ti}^{\rm new}$ becomes negative, it is set to zero. Thus only the energy
deposition higher than the confidential interval $k\cdot D_T$ around the mean 
value is taken into
account decreasing the background influence by such a way. Let us emphasize 
that due to event-by-event background fluctuations, subtracting not only 
the average energy 
deposition but also its dispersion with some factor $k>0$ is necessary (although
this results also in some reduction of a signal itself). The
similar procedure is applying to increase the efficiency of jet reconstruction 
in heavy ion collisions at LHC~\cite{cern,baur}. 

To be specific, we consider the geometry of the 
CMS detector~\cite{cms94} at LHC. The central (``barrel'') part of the CMS 
calorimetric system  covers the pseudo-rapidity region $|\eta| < 1.5$, the 
segmentation of electromagnetic and hadron calorimeters being $\Delta \eta 
\times \Delta \phi = 0.0174 \times 0.0174$ and $\Delta \eta \times \Delta \phi 
= 0.0872 \times 0.0872$ respectively~\cite{cms94}. In order to reproduce 
roughly the experimental conditions (not including real detector effects, but 
just assuming calorimeter hermeticity), we applied (\ref{erecal}) to the 
energy deposition of the generated particles, integrated over the rapidity in 
$N=72$ sectors (according to the number of sectors in the hadron calorimeter) 
covering the full azimuth. Then the energy-energy correlation function 
(\ref{4}) is calculated for the events containing at least one jet with 
$E_T^{\rm jet} > E_T(\rm threshold) = 100$ GeV in the considered kinematical 
region. The final jet energy is defined here as the total transverse energy of 
final particles collected around the direction of a leading particle inside the 
cone $R=\sqrt{\Delta \eta ^2+\Delta \varphi ^2}=0.5$, where $\eta$ and 
$\varphi$ are the pseudo-rapidity and the azimuthal angle respectively. Note
that the estimated event rate in the CMS acceptance, $\sim 10^7$ jets with
$E_T>100$ GeV in a one month LHC run with lead beams~\cite{cern,baur}, will be 
large enough to carefully study the energy-energy correlations over the whole
azimuthal range.  

Before discussing numerical results, let us enlarge upon the role of background
fluctuations for our analysis. The simple estimation on this can be obtained
from the following considerations. The average (over events) energy deposition 
in a calorimeter sector for the mean total particle multiplicity 
$\left< N_p\right> $ and mean particle transverse momentum $\left< p_T\right> $ 
is approximately $\left< E_{Ti}^{\rm thermal}\right> \approx \left< p_T\right> 
\cdot \left< N_p\right> /N$. Then the energy dispersion in a calorimeter
sector can be roughly estimated as $\sigma _T^{\rm thermal} \sim \left< 
p_T\right> \cdot \sqrt{\left< N_p\right> /N}$. For realistic values $N=72$, 
$\left< p_T\right> = 0.55$ GeV/$c$ and $\left< N_p\right> \sim 20000$ 
(corresponds to the charged particle density per unit rapidity 
$dN^{\pm}/dy (y=0) = 5000$ and $\Delta \eta$=3), we get $\left< E_{Ti}^{\rm 
thermal}\right> \sim 150$ GeV and $\sigma _T^{\rm thermal} \sim 10$ GeV. 
Namely these background parameters were fixed for our calculations.
Note that a really measurable energy flux (and correspondingly absolute value 
of its fluctuations) can be essentially less than above values, because of 
limited experimental capabilities to detect low-p$_T$ particles which giving 
bulk of total multiplicity. In particular, in CMS most of charged low-$p_T$ 
particles may be cleared out of the central calorimeters by the strong magnetic 
field. Thus the analysis of quality of extracting jet-like energy-energy 
correlator from background fluctuations based on the detailed simulation of 
detector responses (together with the optimization of jet detection threshold) 
should be performed for each specific experiment.  

The results of numerical simulations are presented on Fig.1 for the cases 
without and with partonic energy loss, two parameterizations of distribution on 
gluon emission angles (\ref{sar}) and (\ref{war}) being used. The important 
issue is the procedure of event-by-event background subtraction (\ref{erecal}) 
(with the factor $k=2$ here) allows the extraction of jet-like energy-energy 
correlation function to be done even in high multiplicity heavy ion events. 
The efficiency of such strategy is closely related with the following. 
Due to the background isotropy, the average (over events) ``thermal'' energy 
deposition in the given calorimeter sector, $\left< E_{Ti}^{\rm thermal}\right>
$, is approximately equal to the average ``thermal'' energy deposition in a 
sector in the given event, $\overline{E_{Ti}^{\rm thermal}( \varphi )}$, at 
the large enough $N$ and numbers of particles in sectors.

The correlation function is sensitive to the medium-induced partonic energy 
loss and angular spectrum of gluon radiation. Three features of 
medium-modified energy-energy correlation function can be noted: \\ 
{\it (1)} Moderate broadening of the near-side jet region $\varphi \la 0.5$
(more pronounced for the ``small-angular'' radiation), which is related with
discussed in the recent literature jet shape modification~\cite{urs03}. \\   
{\it (2)} Significant strengthening in the wide region of azimuthal angles 
around $\varphi \simeq \pi/2$ (more pronounced for the ``broad-angular'' 
radiation) at the relatively small, but hopefully still statistically reliable 
signal values. \\  
{\it (3)} Small additional suppression of back-to-back correlations for 
$\varphi \simeq \pi$ (the analog of acoplanarity~\cite{acop}) as compared to 
the original (in $pp$) suppression due to the initial and final state gluon 
radiation.  

The observation of above modifications demands both the high enough statistics 
($\ga 10^6$ events) and the fine azimuthal resolution of jet position
($\la 0.1$ rad) which, however, are expected to be attainable in calorimetric
measurements at LHC~\cite{cern,baur}. 

Note that for events with the trigger on the 
pair of energetic jets ($E_T^{\rm jet}(1,2) > E_T(\rm threshold) = 100$ GeV) 
the calculated correlator is practically insensitive to the partonic energy 
loss, mainly, because such jets must lose small amount of energy to be visible 
simultaneously. 

\section{Conclusions} 

In summary, at the special selection of events for the analysis (at least one
high-$p_T$ jet) and the procedure of event-by-event background subtraction 
in high multiplicity heavy ion collisions, the transverse energy-energy 
correlator is sensitive to the partonic energy loss and angular spectrum of 
radiated gluons. Medium-modified energy-energy correlation
function manifests significant strengthening in the wide interval of azimuthal 
angles around $\pi/2$, moderate broadening of the near-side jet region  
$\varphi \la 0.5$ and weak additional suppression of back-to-back correlations 
for $\varphi \sim \pi$. Without jet trigger 
this correlation function shows the global structure of transverse energy flux:
the correlator is isotropic for central collisions and for non-central 
collisions it is sensitive to the azimuthal anisotropy of energy flow 
reproducing its Fourier harmonics but with the coefficients squared. We believe 
that such energy-energy correlation analysis may be useful at LHC data 
processing.

\bigskip

{\it Acknowledgments}.  Discussions with Yu.L. Dokshitzer, S.V. Petrushanko, 
I.N. Vardanyan, B.G. Zakharov and G.M. Zinovjev are gratefully acknowledged. 
This work is supported by grant N 04-02-16333 of Russian Foundation for Basic 
Research.

\begin{figure}[hbtp] 
\begin{center} 
\makebox{\epsfig{file=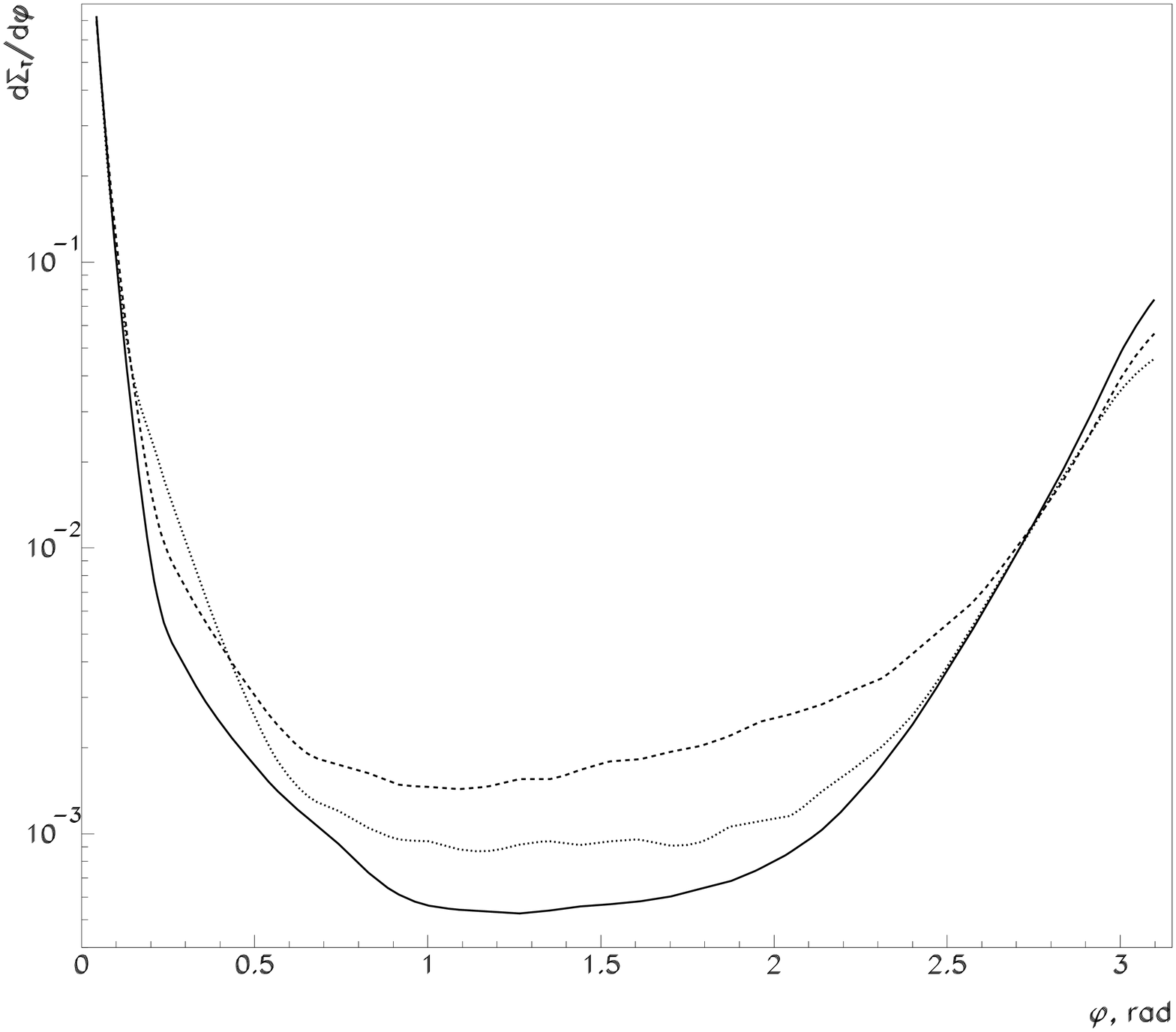, height=170mm}}   
\vskip 1cm 
\caption{The transverse energy-energy correlator $d\Sigma_T/d\varphi$ 
as a function of azimuthal angle $\varphi$
without (solid histogram) and with medium-induced partonic 
energy loss for the ``small-angular'' (\ref{sar}) (dotted histogram) and the 
``broad-angular'' (\ref{war}) (dashed histogram) parameterizations of emitted 
gluon spectrum in central Pb$-$Pb collisions. Applied kinematical cuts are
described in the text. }  
\end{center}
\end{figure}

\end{document}